\documentclass[lettersize,journal]{IEEEtran}
\IEEEoverridecommandlockouts
\usepackage{cite}
\usepackage{amsmath,amsfonts,amssymb}
\usepackage{array}
\usepackage[caption=false,font=normalsize,labelfont=sf,textfont=sf]{subfig}
\usepackage{textcomp}
\usepackage{stfloats}
\usepackage{url}
\usepackage{verbatim}
\usepackage{graphicx}
\usepackage{cite}
\usepackage{amsmath}
\usepackage{textcomp}
\usepackage{xcolor}
\usepackage{subfig}
\usepackage{hyperref}
\usepackage[procnumbered,ruled,vlined]{algorithm2e}
\usepackage[normalem]{ulem}
\usepackage{multirow}
\usepackage{xfrac}
\usepackage{inputenc}
\usepackage{color}
\usepackage{booktabs}
\usepackage{makecell}
\usepackage{enumitem}

\title{FlexSpec: Frozen Drafts Meet Evolving Targets in Edge-Cloud Collaborative LLM Speculative Decoding}

\author{
\IEEEauthorblockN{Yuchen Li, Rui Kong, Zhonghao Lyu, Qiyang Li, Xinran Chen, Hengyi Cai, Lingyong Yan, Shuaiqiang Wang, Jiashu Zhao, Guangxu Zhu, Linghe Kong,~\IEEEmembership{Fellow, IEEE}, Guihai Chen,~\IEEEmembership{Fellow, IEEE}, Haoyi Xiong, Dawei Yin}
\thanks{Y. Li, R. Kong, Q. Li, X. Chen, H. Cai, L. Yan, S. Wang, H. Xiong, and D. Yin are with Baidu Inc., Beijing, China (e-mail: \{yuchenli1230, monster119120, qiyangli878, fantastique0910, hengyi1995, lingyongy, shqiang.wang\}@gmail.com; haoyi.xiong.fr@ieee.org; yindawei@acm.org).}
\thanks{Y. Li, L. Kong, and G. Chen are with the School of Computer Science, Shanghai Jiao Tong University, Shanghai, China (e-mail: \{yuchenli, linghe.kong, chen-gh\}@sjtu.edu.cn).}
\thanks{Z. Lyu is with the Department
of Information Science and Engineering, KTH Royal Institute of Technology, Stockholm, Sweden (e-mail: lzhon@kth.se). {\it (Corresponding author: Z. Lyu)}}
\thanks{J. Zhao is with the Department of Physics and Computer Science, Wilfrid University (e-mail: jzhao@wlu.ca).}
\thanks{G. Zhu is with the Shenzhen Research Institute of Big Data, Shenzhen, China (e-mail: gxzhu@sribd.cn).}
}

\begin{document}

\maketitle

\begin{abstract}
Deploying large language models (LLMs) in mobile and edge computing environments is constrained by limited on-device resources, scarce wireless bandwidth, and frequent model evolution. 
Although edge-cloud collaborative inference with speculative decoding (SD) can reduce end-to-end latency by executing a lightweight draft model at the edge and verifying it with a cloud-side target model, existing frameworks fundamentally rely on tight coupling between the two models.
Consequently, repeated model synchronization introduces excessive communication overhead, increasing end-to-end latency, and ultimately limiting the scalability of SD in edge environments.
To address these limitations, we propose FlexSpec, a communication-efficient collaborative inference framework tailored for evolving edge-cloud systems. The core design of FlexSpec is a shared-backbone architecture that allows a single and static edge-side draft model to remain compatible with a large family of evolving cloud-side target models. By decoupling edge deployment from cloud-side model updates, FlexSpec eliminates the need for edge-side retraining or repeated model downloads, substantially reducing communication and maintenance costs. Furthermore, to accommodate time-varying wireless conditions and heterogeneous device constraints, we develop a channel-aware adaptive speculation mechanism that dynamically adjusts the speculative draft length based on real-time channel state information and device energy budgets. Extensive experiments demonstrate that FlexSpec achieves superior performance compared to conventional SD approaches in terms of inference efficiency.
\end{abstract}


\section{Introduction}
\label{sec:intro}

\IEEEPARstart{L}{arge} language models (LLMs), exemplified by ChatGPT \cite{achiam2023gpt}, LLaMA \cite{touvron2023llama2}, and Gemini \cite{team2023gemini}, have emerged as a critical engine for intelligent agents, ranging from mobile personal assistants to real-time translation devices. The capability of these models largely determines the quality of user experience and the successful execution of complex reasoning tasks~\cite{li2025towards,li2025m,xiong2024search,li2025s}. Although many studies have focused on compressing these models on edge devices via techniques like quantization and pruning \cite{liu2024mobilellm}, individual on-device inference remains constrained by factors such as limited battery life, thermal throttling, and insufficient memory bandwidth \cite{xu2021edge}.

Edge-cloud collaborative inference addresses this limitation by offloading computationally intensive sub-tasks to powerful cloud servers while retaining lightweight processing at the edge, thereby expanding the inference capability and improving response accuracy \cite{liu2025edge}. A critical challenge in this collaborative paradigm lies in maintaining real-time performance without sacrificing accuracy, especially in highly dynamic mobile network environments. Delays between the user query and the generated response can render the output frustratingly slow, thereby compromising the interactivity and reliability of the application. This issue becomes even more pronounced in autoregressive generation, where token-by-token transmission introduces significant communication latency. Furthermore, the amount of information exchanged between the edge and cloud is often substantial. Consequently, efficient inference acceleration methods are essential to reduce communication overhead and decision latency~\cite{10791415}.

To mitigate the latency of autoregressive generation, speculative decoding (SD) has been widely adopted as a key inference acceleration technique~\cite{leviathan2023fast}.
In a typical edge-cloud SD paradigm, the edge device rapidly generates a sequence of speculative ``draft'' tokens, which are transmitted to a powerful cloud server for parallel verification. While recent deep learning-based SD methods, such as EAGLE~\cite{li2024eagle} and Medusa~\cite{cai2024medusa}, have demonstrated promising latency reduction, their deployment in practical edge-cloud systems faces several fundamental challenges.
Specifically, existing SD frameworks rely on a tight coupling between the draft and target models, while cloud-side target models are frequently updated via parameter-efficient fine-tuning (PEFT) \cite{10791415}, leading to an ``\emph{update storm}''
characterized by excessive communication overhead for maintaining model consistency over bandwidth-constrained wireless links. Moreover, avoiding model synchronization by keeping the edge draft model static introduces severe distribution shifts as the target model evolves, resulting in
a sharp degradation in token acceptance rates, commonly referred to as ``\emph{performance collapse}''.
Finally, speculative decoding performance is highly sensitive to wireless channel dynamics, where time-varying uplink latency can significantly undermine the latency gains of fixed-stride speculative execution in mobile environments. Taken together, frequent model evolution, distribution shift between draft and target models, and time-varying wireless latency fundamentally limit the applicability of existing speculative decoding frameworks in practical edge-cloud deployments.

Motivated by these challenges, our goal is to realize \textit{version-agnostic} speculative decoding among edge-cloud systems, i.e., accelerating diverse, evolving cloud target models through a single static edge model. Once we construct this robust drafting framework, it can enable acceleration across multiple fine-tuned versions using a single draft model with significantly reduced maintenance costs. Moreover, this decoupled framework generalizes well to unseen target distributions, unlike existing systems that must be retrained from scratch. As a result, both deployment and communication costs are substantially reduced.

Specifically, inspired by the success of foundation models (FMs) in modality-agnostic processing,  we propose \textbf{FlexSpec}, a communication-efficient edge-cloud collaborative speculative decoding framework for evolving large language models. The key idea of FlexSpec is to fundamentally decouple the lifecycle of the edge-side draft model from that of the cloud-side target model, thereby eliminating the need for frequent over-the-air model synchronization in bandwidth-constrained wireless environments. Moreover, FlexSpec explicitly accounts for the stochastic nature of wireless channels by adapting the speculative decoding behavior to time-varying network conditions. By jointly considering model evolution and wireless dynamics within a unified framework, FlexSpec enables scalable, robust, and low-latency LLM inference in edge systems.
The main contributions of this paper are summarized as follows:
\begin{itemize}
    \item We propose FlexSpec, a novel edge-cloud collaborative speculative decoding framework that enables \emph{version-agnostic} inference for evolving large language models. By introducing a shared frozen anchor backbone between the edge draft model and the cloud target model, FlexSpec fundamentally decouples their lifecycles and eliminates the need for frequent draft model synchronization, effectively addressing the ``update storm'' problem in wireless edge networks.
    \item We develop a channel-adaptive drafting policy for speculative decoding under edge environments. By modeling the interaction between speculative stride length, token acceptance rate, and communication rate, we derive a theoretically grounded formulation for the drafting length selection, enabling dynamic adaptation to time-varying channel conditions and maximizing effective token generation throughput.
    \item We conduct extensive experiments on a large-scale GPU cluster (up to 32 NVIDIA A800 GPUs) across diverse reasoning and coding tasks (e.g., GSM8K and HumanEval). The results demonstrate that FlexSpec consistently improves end-to-end generation efficiency and delivers robust performance across heterogeneous wireless networks. In particular, by adaptively adjusting the speculative stride to time-varying channel conditions, FlexSpec outperforms state-of-the-art speculative decoding methods in communication-constrained regimes. Moreover, by decoupling a frozen edge draft model from frequently updated cloud-side targets, FlexSpec largely avoids recurrent draft-model synchronization, thereby substantially reducing update-related communication overhead.
\end{itemize}

The rest of this paper is organized as follows. Section II presents the related work. Section III introduces the system model and problem definition for edge-cloud speculative decoding. Section IV details the network architecture and the adaptive policy of the proposed FlexSpec. Section V shows experimental results to verify the performance of FlexSpec. Section VI concludes this paper.

\section{Related Work}
\label{sec:related_work}
In this section, we review and discuss the related works from the perspectives of \emph{SD for LLM Acceleration}, \emph{Edge-Cloud Collaborative Inference}, and \emph{Wireless-Aware Mobile AI}.

\subsection{SD for LLM Acceleration}
SD has emerged as a pivotal paradigm to alleviate the memory-bound latency bottleneck inherent in the autoregressive generation of LLMs. The core premise of SD is to trade abundant parallel compute capability for reduced sequential memory access by employing a ``draft-then-verify'' strategy.

In the foundational works, Leviathan et al. \cite{leviathan2023fast} and Chen et al. \cite{chen2023accelerating} formalized the speculative execution framework, demonstrating that a small draft model can generate candidate tokens which are then verified in parallel by the target model, guaranteeing mathematically identical outputs to the target model. Building on this, recent advancements have focused on improving draft quality and acceptance rates. Medusa \cite{cai2024medusa} introduces multiple decoding heads on top of the frozen original model to predict multiple future tokens simultaneously, eliminating the need for a separate draft model. EAGLE \cite{li2024eagle} and EAGLE-2 \cite{li2024eagle2} further advance this by utilizing layer-wise feature extrapolation, processing input features at the draft layer to generate contextual drafts with higher accuracy. Beyond model-based approaches, non-model drafting techniques such as n-gram matching \cite{yang2023inference} and retrieval-based augmentation \cite{he2023rest} utilize statistical patterns or external databases to propose tokens without heavy computation.

However, these existing SD frameworks operate under the assumption of a static environment where the draft and target models are tightly coupled or structurally identical. In edge scenarios, cloud target models evolve frequently via PEFT \cite{mangrulkar2022peft}. Maintaining the strict coupling imposed by methods such as EAGLE or Medusa would require frequent synchronization of the edge-side draft model over wireless links with the cloud target model, thereby incurring prohibitive communication overhead. FlexSpec differentiates itself by structurally decoupling the draft and target lifecycles, enabling a single static edge model to serve evolving cloud targets without frequent synchronization.

\subsection{Edge-Cloud Collaborative Inference}
Edge-cloud collaborative inference bridges the gap between the limited resource capacity of mobile devices and the massive computational demands of modern deep neural networks (DNNs) \cite{10388062}. This field generally investigates how to optimally partition computation between the edge and the cloud to balance latency, energy consumption, and privacy.

Early works, such as Neurosurgeon \cite{kang2017neurosurgeon} and technologies explored by Matsubara et al. \cite{matsubara2019split}, focused on identifying the optimal split point within a DNN to minimize end-to-end latency. With the rise of LLMs, decentralized inference frameworks (e.g., \cite{borzunov2023petals,ryabinin2023swarm,11301737}) have been proposed to distribute transformer blocks across heterogeneous devices over the internet. Researchers have proposed aggressive quantization techniques \cite{frantar2022gptq, lin2023awq} and activation compression methods \cite{xie2023distri} to alleviate the communication bottleneck caused by transmitting large intermediate activation tensors.

Despite these advances, standard split inference approaches often suffer from high transmission latency when applied to generative tasks, as transmitting high-dimensional activations for every token is communication-intensive. Furthermore, they largely overlook the version consistency issue in a production environment where the cloud model updates daily. FlexSpec advances this field by transmitting lightweight token indices instead of heavy activations and by introducing an anchor-based alignment mechanism that makes the edge-cloud collaboration robust to model version mismatches.

\subsection{Wireless-Aware Mobile AI}
The deployment of AI applications in mobile computing environments necessitates a rigorous consideration of wireless network dynamics. Unlike stable wired connections, dynamic wireless channels with fading and noise significantly impact the Quality of Experience (QoE) for real-time AI applications. To tackle such issues,  Park et al. \cite{park2019wireless} and Zhu et al. \cite{zhu2020toward} laid the groundwork for wireless edge intelligence by optimizing the trade-off between computation offloading and communication channel quality. In the specific context of NLP, dynamic inference mechanisms like DeeBERT \cite{xin2020deebert} and SPINN \cite{laskaridis2020spinn} allow models to exit early based on confidence thresholds to save energy. More recently, DSSD \cite{chen2023fos} attempted to adapt speculative decoding for wireless setups but relied on fixed strategies that do not fully exploit real-time channel state information.

FlexSpec fills the gap between wireless networking theory and LLM inference systems. Unlike prior works that treat the generation length $K$ as a fixed hyperparameter, FlexSpec models the drafting stride as a dynamic variable dependent on the channel conditions. By applying channel-aware adaptation to the generative process, FlexSpec ensures that the system maintains high throughput even under the volatile network conditions characteristic of mobile scenarios.

\section{System Model and Problem Definition}
\label{sec:motivation}

In this section, we first present the system model for edge-cloud collaborative inference. Then, based on preliminary experiments using real-world datasets and network traces, we identify three critical research challenges that hinder the
practical deployment of existing speculative decoding frameworks in mobile environments: the prohibitive bandwidth cost of model synchronization, the performance collapse induced by distribution shifts, and the sensitivity to wireless network volatility.

\subsection{System Model: Edge-Cloud Collaborative Inference}
We consider a collaborative inference system consisting of an edge device $\mathcal{E}$ and a cloud server $\mathcal{C}$. The cloud hosts a high-capacity \emph{target model} $\mathcal{M}_t(\cdot;\theta_t)$ (e.g., Llama-2-70B) deployed on GPU clusters (e.g., NVIDIA A800), where $\theta_t \in \mathbb{R}^{N_t}$ denotes its parameters. To maintain competitive performance on evolving downstream tasks, the service provider periodically updates $\mathcal{M}_t$ via PEFT or full-parameter training on an evolving task dataset $\mathcal{D}_{\text{task}}$. Consequently, the cloud model evolves over time as a sequence $\{\mathcal{M}_t^{(0)},\mathcal{M}_t^{(1)},\dots,\mathcal{M}_t^{(s)}\}$.

On the edge, the client hosts a lightweight \emph{draft model} $\mathcal{M}_d(\cdot;\theta_d)$ with $\theta_d \in \mathbb{R}^{N_d}$ and $N_d \ll N_t$, whose size is typically comparable to only a small fraction of $\mathcal{M}_t$ (e.g., on the order of a single target-model layer). Due to stringent constraints on storage, battery, and computation capabilities, the edge device cannot afford frequent downloads of large model weights. Therefore, it is kept \emph{static} (frozen) across target-model updates, i.e., $\mathcal{M}_d^{(s)}=\mathcal{M}_d^{(0)}, \forall s$.

The edge and cloud communicate over a stochastic wireless link. In each speculative decoding step, the edge drafts a block of tokens $x_{\text{draft}}$ and transmits them to the cloud, while the cloud verifies the drafted block using $\mathcal{M}_t$ and returns the verification outcomes  $x_{\text{verified}}$. 

Given an input prompt $X$, the system generates an output sequence $Y=\{y_1,\dots,y_O\}$ with length $O$. Under speculative decoding with draft length $K$, the latency of one decoding step $n$ is modeled as
\begin{equation}
    T_{\text{step}}(K, R_n)=T_{\text{edge}}(K)+T_{\text{up}}(K, R_n)+T_{\text{cloud}}(K)+T_{\text{down}},
\label{eq:latency_model}
\end{equation}
where $T_{\text{edge}}(K)$ is the edge computation time to draft $K$ tokens, $T_{\text{up}}(K, R_n)$ is the uplink transmission delay, $T_{\text{cloud}}(K)$ is the cloud-side verification time, and $T_{\text{down}}$ is the downlink latency to deliver verification results/feedback. Moreover, $T_{\text{up}}(K, R_n)$ is further expressed as the ratio between the size of uplink transmission overhead $B_{\text{up}}(K)$ and the achievable uplink rate $R_n$ at step $n$, i.e., $T_{\text{up}}(K, R_n)=B_{\text{up}}(K)/R_n$. 

\begin{table*}[h]
\centering
\caption{Estimated Latency for Synchronizing Draft Models over Wireless Networks. \\ Frequent updates render standard SD infeasible.}
\label{tab:sync_overhead}
\begin{tabular}{l|c|c|c}
\toprule
\textbf{Network Type} & \textbf{Bandwidth} & \textbf{Synchronization Time (one User)} & \textbf{Scalability (1k Users)} \\
\midrule
Public WiFi & 10 Mbps & $\approx$ 48 min & \textbf{Collapse} \\
4G LTE & 50 Mbps & $\approx$ 9.5 min & High Congestion \\
5G Mid-Band & 300 Mbps & $\approx$ 1.6 min & Moderate Load \\
\bottomrule
\end{tabular}
\end{table*}


\subsection{Challenge 1: The ``Update Storm'' in Wireless Networks}
The first major obstacle is the conflict between the need for model freshness and limited wireless bandwidth. In standard SD paradigms, the draft model $\mathcal{M}_d$ is required to approximate the distribution of the target model $\mathcal{M}_t$. Consequently, whenever $\mathcal{M}_t$ is updated to a new version $\mathcal{M}_t'$ (e.g., fine-tuned for a medical application), the edge-side $\mathcal{M}_d$ must theoretically be retrained and synchronized to maintain high token acceptance rates $\hat \gamma$.

We define this phenomenon as the ``update storm.'' To quantify its impact, we conducted a preliminary analysis of the transmission overhead required to synchronize a lightweight draft model (approximately
3.2~GB) over typical mobile networks.

As shown in Table \ref{tab:sync_overhead}, downloading even a compressed draft model takes nearly 10 minutes on a 4G network. For a mobile app with daily model updates, this imposes petabytes of traffic on the cellular infrastructure and rapidly drains the user's data plan and battery. This motivates our design of a \textit{static} draft model that does not require synchronization.

\subsection{Challenge 2: Distribution Shift and Performance Collapse}
If we forgo model synchronization to save bandwidth (i.e., keeping $\mathcal{M}_d$ fixed while $\mathcal{M}_t$ evolves), we encounter the second challenge: distribution shift. The discrepancy between the static draft model's output distribution $P_d$ and the evolving target model's distribution $P_t'$ grows significantly after fine-tuning.

To verify this, we performed a preliminary experiment using a generic Llama-2-7B as the frozen draft model. We measured the token acceptance rate  against three versions of the 70B target model: the Base version, a Math-tuned version (GSM8K), and a Code-tuned version (HumanEval).

\begin{table}[h]
\centering
\caption{Impact of Target Model Evolution on Fixed Draft Model Performance. The acceptance rate drops drastically without alignment.}
\label{tab:dist_shift}
\begin{tabular}{l|c|c}
\toprule
\textbf{Target Model Version} & \textbf{Domain} & \textbf{Acceptance Rate} \\
\midrule
Llama-2-70B-Base & General & \textbf{0.72} \\
Llama-2-70B-Math (LoRA) & Mathematics & 0.45 ($\downarrow$ 37.5\%) \\
Llama-2-70B-Code (Full) & Programming & \textbf{0.18 ($\downarrow$ 75.0\%)} \\
\bottomrule
\end{tabular}
\end{table}

The results in Table \ref{tab:dist_shift} reveal a severe ``performance collapse.'' When the target model evolves to the coding domain, the acceptance rate of the generic draft model plummets to 0.18. At this low acceptance rate, the overhead of speculative decoding (running the draft model and transmitting tokens) outweighs the benefits, often resulting in higher latency than standard autoregressive decoding. This necessitates a mechanism like FlexSpec's \textit{anchor-based alignment} to bridge the gap between static and evolving representations.

\subsection{Challenge 3: Wireless Latency Sensitivity}
The third challenge lies in the coupling between speculative length $K$ and network conditions.
Most existing works adopt a fixed speculative stride $K$ (e.g., $K=5$). However, our network trace analysis indicates that a fixed $K$ can be suboptimal in mobile environments.
In weak-signal scenarios (e.g., elevators or subways with signal-to-noise ratio (SNR) $<5$~dB), the uplink transmission term $T_{\text{up}}(K, R_n)$ in \eqref{eq:latency_model} dominates the end-to-end latency.
For instance, transmitting five tokens may incur approximately 200~ms of uplink delay, whereas the corresponding verification gain is on the order of 50~ms. Conversely, under strong 5G channel conditions, a small fixed $K$ fails to fully utilize the available bandwidth. This mismatch creates a fundamental dilemma: a static speculation strategy
cannot simultaneously meet the latency requirements across heterogeneous mobility and channel conditions, thereby highlighting the need for the \textit{channel-aware adaptive speculation} mechanism proposed in this paper.

\subsection{Problem Definition}
Based on \eqref{eq:latency_model}, we aim to design an edge-cloud SD framework that remains effective in evolving edge-cloud systems. Specifically, given a sequence of cloud-side target models $\{\mathcal{M}_t^{(s)}\}$ updated over time, we seek a model-coupling mechanism that allows the edge draft model $\mathcal{M}_d$ to stay \emph{static} (i.e., without edge-side retraining or repeated model downloads) while preserving high token acceptance when cooperating with each $\mathcal{M}_t^{(s)}$.

Meanwhile, under time-varying wireless conditions, we aim to seek a policy on $K$ and a model-coupling mechanism between $\mathcal{M}_d$ and $\mathcal{M}_t$ to maximize the effective token generation rate (ETGR), i.e., the expected number of accepted tokens per unit time. Let $\tau$ denote the number of tokens accepted in one draft-and-verify round with draft length $K$, then the ETGR is characterized by the ratio between the expected number of accepted tokens
$\tau$ resulting from a speculative draft of length $K$, i.e., $\mathbb{E}[\tau\mid K]$, and the per-round latency in \eqref{eq:latency_model}, i.e., 
\begin{equation}
    \mathrm{ETGR}(K)
    \triangleq
    \frac{\mathbb{E}\!\left[\tau\mid K \right]}
    {\mathbb{E}\!\left[T_{\text{step}}(K, R_n)\right]},
    \label{eq:etgr_def}
\end{equation}
which is equivalent to minimizing the expected latency per accepted token, i.e.,
$\mathbb{E}[T_{\mathrm{step}}(K)]/\mathbb{E}[\tau\mid K]$.
In summary, we aim to learn an adaptive policy on $K$ and a model-coupling mechanism between $\mathcal{M}_d$ and $\mathcal{M}_t$ that maximize ETGR.

\section{Methodology: FlexSpec}
\label{sec:methodology}

In this section, we detail the FlexSpec framework to tackle the aforementioned challenges. Our approach consists of three integrated components: (1) A \textit{structural decoupling architecture} that enables a static edge draft model to serve evolving cloud targets; (2) A \textit{channel-aware speculation policy} that optimizes the drafting length $K$ based on real-time channel conditions; and (3) A \textit{semantic synchronization protocol} that minimizes transmission overhead.




\subsection{Architecture: Anchor-Based Feature Alignment}
\label{subsec:architecture}

\begin{figure*}[t]
    \centering
    \includegraphics[width=0.9\textwidth]{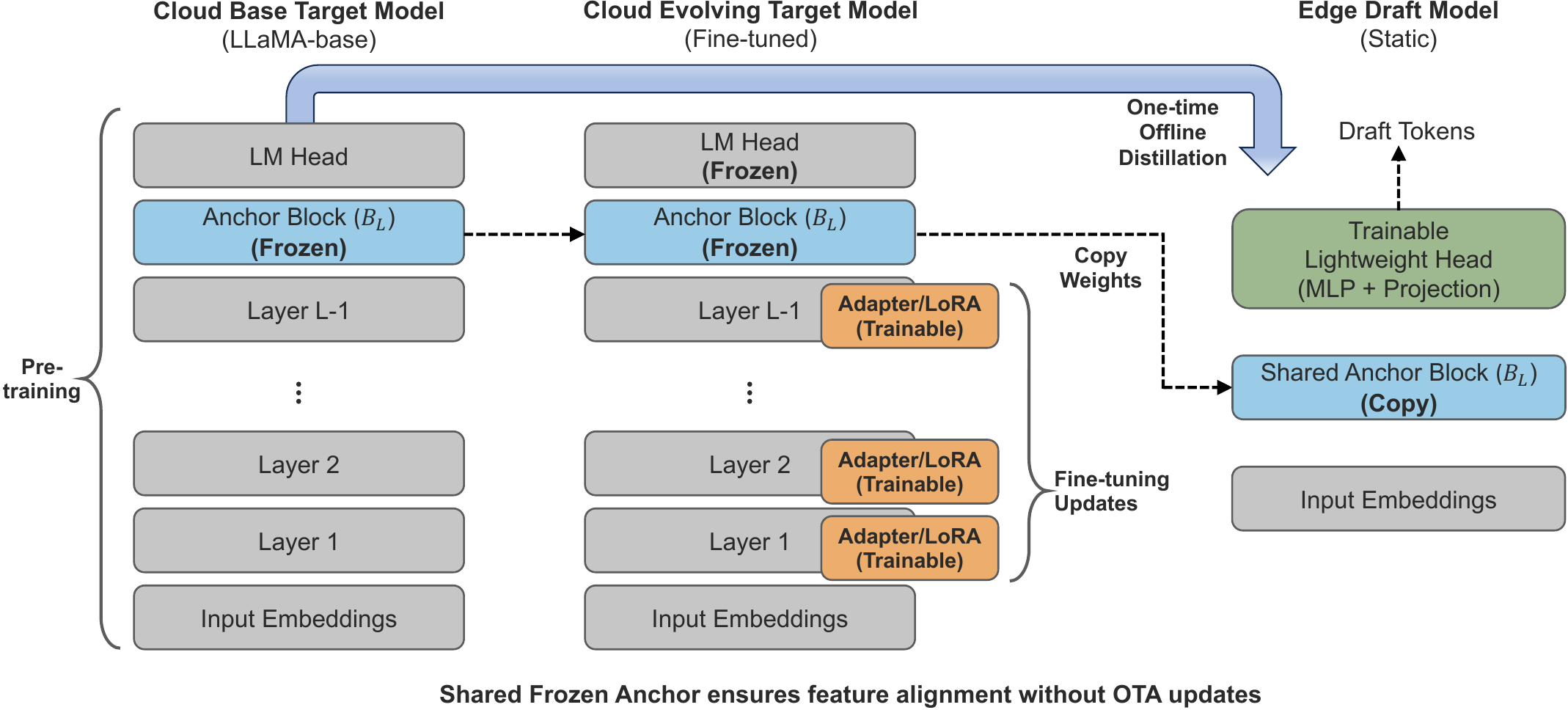}
    \caption{The FlexSpec training architecture: A shared frozen anchor block ($B_L$) ensures feature space alignment between the static edge draft model and the evolving cloud target model. The lightweight head is trained via one-time offline distillation.}
    \label{fig:training}
\end{figure*}

To satisfy the constraint that $\theta_d$ remains static while $\theta_t$ evolves, we propose the anchor-based alignment architecture. Standard speculative decoding fails when the distribution $P_d(y|x)$ diverges from $P_t(y|x)$ due to cloud-side fine-tuning. We hypothesize that while high-level semantic instructions change during fine-tuning, the fundamental token representation logic remains stable in the backbone \cite{merchant2023finetuning}.

\subsubsection{Structural Definition}
We denote by $\mathcal{M}_{\text{base}}\triangleq \mathcal{M}_t^{(0)}$ the cloud target model before task-specific fine-tuning. Following the architecture in Fig.~\ref{fig:training}, we decompose $\mathcal{M}_{\text{base}}$ into a feature extractor and a head as
\begin{equation}
\mathcal{M}_{\text{base}}(x) = \Psi_{\text{base}}\!\big(\Phi_{\text{base}}(x)\big),
\end{equation}
where $\Phi_{\text{base}}$ consists of the bottom transformer stack (layers $1$ to $L\!-\!1$) together with the input embedding module, and $\Psi_{\text{base}}$ denotes the head that contains the last transformer block (layer $L$) and the LM head. We refer to the last transformer block inside $\Psi_{\text{base}}$ as the \emph{anchor block}, denoted by $\mathcal{B}_L^{\text{base}}$.

Based on this decomposition, the edge draft model $\mathcal{M}_d$ is constructed by (i) copying the anchor block from the base head and keeping it frozen, and (ii) attaching a lightweight trainable head for token prediction. Specifically,
\begin{equation}
\mathcal{M}_d(x)=\mathcal{H}_{\text{small}}\!\big(\mathcal{B}_{\text{shared}}(x)\big),
\end{equation}
where $\mathcal{B}_{\text{shared}}\triangleq \mathcal{B}_L^{\text{base}}$ is a frozen copy of the anchor block, and $\mathcal{H}_{\text{small}}$ is a lightweight trainable projection head consisting of a two-layer multilayer perceptron (MLP) followed by a vocabulary projection matrix. In this way, $\mathcal{M}_d$ reuses the representation space induced by the base head while remaining sufficiently compact for edge deployment.

During cloud-side fine-tuning, we adopt a PEFT-style parameterization for the evolving target model $\mathcal{M}_t(\cdot;\theta_t)$ by decomposing its parameters as $\theta_t=\{\theta_t^{\text{backbone}},\theta_t^{\text{adapters}}\}$. Here, $\theta_t^{\text{backbone}}$ denotes the original pre-trained weights of $\mathcal{M}_{\text{base}}$ (i.e., the parameters of $\Phi_{\text{base}}$ and $\Psi_{\text{base}}$, including the anchor block and the LM head), while $\theta_t^{\text{adapters}}$ denotes the additional PEFT parameters (e.g., LoRA/adapter modules) injected into the transformer layers. We enforce a backbone-freezing constraint by keeping $\theta_t^{\text{backbone}}$ frozen and only updating $\theta_t^{\text{adapters}}$ during fine-tuning. Consequently, the head $\Psi_{\text{base}}$ (layer $L$ and LM head) remains invariant, which stabilizes the feature manifold seen by the anchor block and preserves the feature compatibility that the static edge draft $\mathcal{M}_d$ relies on for accurate drafting.

\subsubsection{Generalist Training Objective}
To train the static edge components ($\mathcal{H}_{small}$), we employ a multi-objective loss function combining feature regression and knowledge distillation (KD) \cite{hinton2015distilling}. The training is performed \textbf{once} on a general corpus (e.g., RedPajama \cite{weber2024redpajama}) using the base model $\mathcal{M}_{base}$ as the teacher. Specifically, the loss functions are shown as follows.
\begin{itemize}
    \item \textbf{Feature Regression Loss ($\mathcal{L}_{feat}$):} $\mathcal{L}_{feat}$ ensures the draft hidden states $h_d$ align with the target hidden states $h_t$, i.e.,
\begin{equation}\label{loss1}
    \mathcal{L}_{feat} = \frac{1}{\omega \cdot \pi} \sum_{i=1}^{\omega} \sum_{j=1}^{\pi} \| W_p \cdot h_{d}^{(i,j)} - h_{t}^{(i,j)} \|_2^2,
\end{equation}
where $W_p$ is a learnable projection matrix to match dimensions \cite{jiao2020tinybert}, and $\omega$ is the batch size and $\pi$ is the sequence length.
\item \textbf{Soft-Target KD Loss ($\mathcal{L}_{KD}$):} It minimizes the KL divergence between the token distributions \cite{gu2023knowledge}, i.e., 
\begin{equation}\label{loss2}
    \mathcal{L}_{KD} = \mathcal{T}^2 D_{KL} \left( \sigma\left(\frac{z_t}{\mathcal{T}}\right) \bigg\| \sigma\left(\frac{z_d}{\mathcal{T}}\right) \right),
\end{equation}
where $z_t, z_d$ are logits, $\sigma$ is the Softmax activation function, and $\mathcal{T}$ is the temperature.
\end{itemize}

Then the total objective is the combination of \eqref{loss1}  and \eqref{loss2}, i.e., $\mathcal{L} = \lambda_1 \mathcal{L}_{feat} + \lambda_2 \mathcal{L}_{KD}$.
Algorithm \ref{alg:training} summarizes this one-time offline training process.

\begin{algorithm}[t]
\caption{FlexSpec offline draft model training}
\label{alg:training}
\SetKwInOut{KwIn}{Input}
\SetKwInOut{KwOut}{Output}

\KwIn{Base target model $\mathcal{M}_{\text{base}}$, dataset $\mathcal{D}_{\text{pretrain}}$}
\KwOut{Static draft model $\mathcal{M}_d = \{\mathcal{B}_{\text{shared}}, \mathcal{H}_{\text{small}}\}$}
\BlankLine

\tcp{Step 1: initialization and structural freezing}
Initialize lightweight head $\mathcal{H}_{\text{small}}$ randomly\;
\tcp{Copy anchor block}
$\mathcal{B}_{\text{shared}} \leftarrow \mathcal{M}_{\text{base}}.\text{block}[-1]$\;
Freeze parameters of $\mathcal{B}_{\text{shared}}$ and $\mathcal{M}_{\text{base}}$\;
\BlankLine

\tcp{Step 2: alignment training loop}
\For{\text{each batch} $(x, y)$ \textbf{in} $\mathcal{D}_{\text{pretrain}}$}{
    \tcp{Teacher forward pass (base model)}
    $h_t, z_t \leftarrow \mathcal{M}_{\text{base}}(x)$\;
    
    \tcp{Student forward pass (draft model)}
    $h_d, z_d \leftarrow \mathcal{H}_{\text{small}}(\mathcal{B}_{\text{shared}}(x))$\;
    
    \tcp{Multi-objective loss computation}
    $\dots$\;
    $\mathcal{L}_{\text{total}} \leftarrow \lambda_1 \mathcal{L}_{\text{feat}} + \lambda_2 \mathcal{L}_{\text{KD}}$\;
    
    Update $\mathcal{H}_{\text{small}}$ via optimizer($\nabla \mathcal{L}_{\text{total}}$)\;
}
\Return $\mathcal{M}_d$
\end{algorithm}

\subsection{Channel-Aware Adaptive Speculation}
\label{subsec:adaptive_policy}

\begin{figure*}[t]
    \centering
    \includegraphics[width=0.9\textwidth]{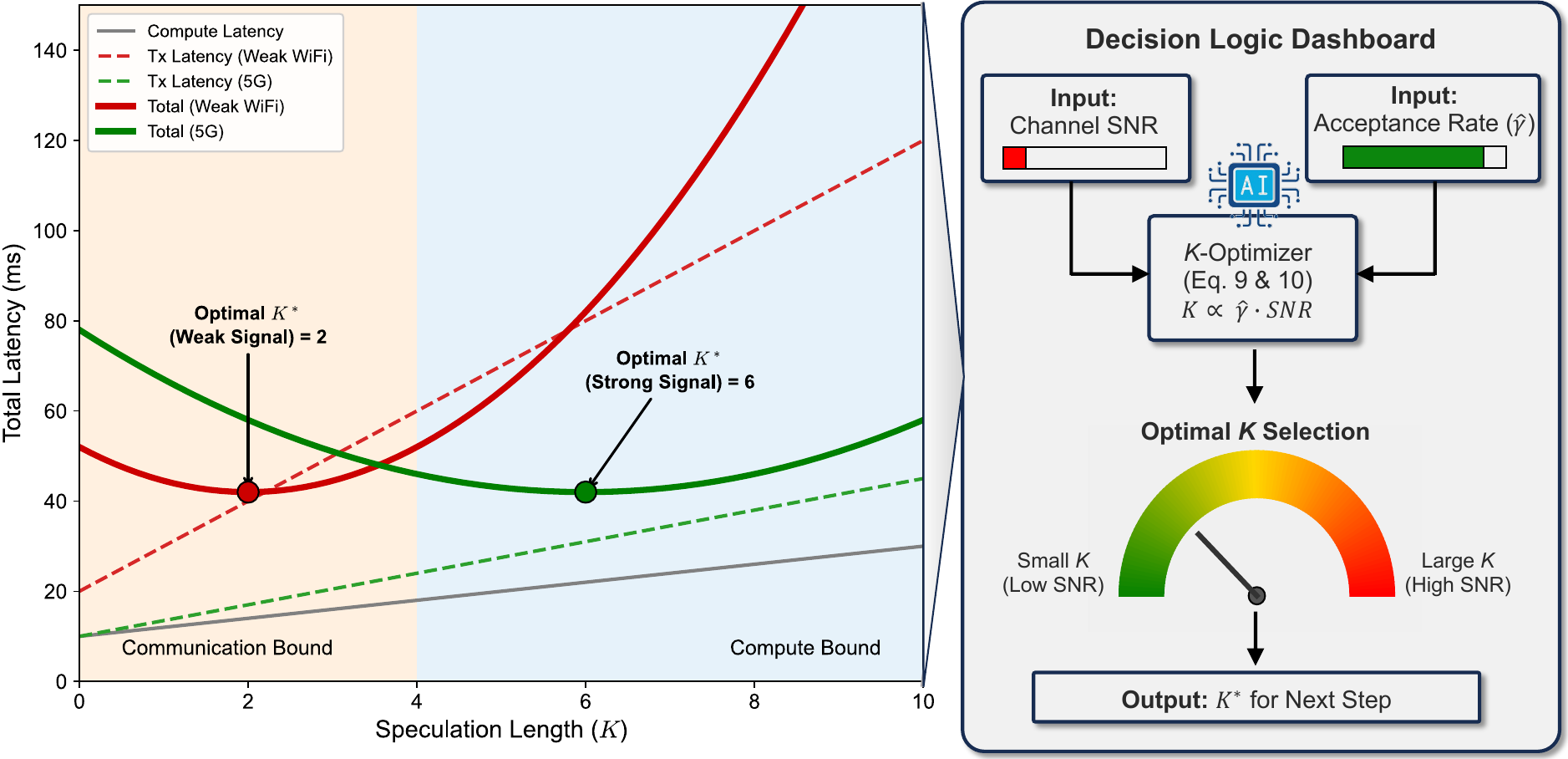}
    \caption{Channel-aware adaptive speculation mechanism: The left plot illustrates the trade-off between communication and compute latency under different signal strengths. The optimal draft length $K^*$ shifts from 2 (Weak Signal) to 6 (Strong Signal).}
    \label{fig:adaptive_k}
\end{figure*}

In mobile computing environments, the end-to-end latency is strictly coupled with channel dynamics. A static draft length $K$ fails to adapt to the trade-off between transmission overhead and speculative gain. To address this, we formulate the optimal $K^*$ selection as a throughput maximization problem, explicitly accounting for propagation delays and variable cloud verification costs.

\subsubsection{Refined Latency Modeling}
The total time consumption of a single speculative decoding step with stride
$K$, denoted by $T_{\text{step}}(K, R_n)$ in~\eqref{eq:latency_model}, consists of four
components: edge-side computation, uplink transmission, cloud-side verification,
and downlink feedback.
Unlike prior works that assume a constant cloud verification cost, we model the
cloud-side latency as an affine function of $K$, which captures the memory
bandwidth overhead incurred by loading $K$ newly generated tokens and their
associated key-value (KV) caches.
Accordingly, the refined latency model is given by
\begin{equation} \label{eq:refined_latency}
    \hat T_{\text{step}}(K, R_n)
    = \hat T_{\text{edge}}(K) + \hat T_{\text{up}}(K, R_n) + \hat T_{\text{cloud}}(K) + T_{\text{down}} .
\end{equation}

The uplink transmission latency explicitly accounts for both the propagation
delay $T_{\text{prop}}$ (equal to half of the round-trip time) and the data
transmission delay determined by the communication overhead $B_{\text{up}}(K) \triangleq K \cdot b + O_{\text{header}}$ and instantaneous communication rate $R_n$,
i.e.,
\begin{equation}
    \hat T_{\text{up}}(K, R_n)
    = T_{\text{prop}} + \frac{K \cdot b + O_{\text{header}}}{R_n},
\end{equation}
where each token is encoded as an integer index of $b$ bits and
$O_{\text{header}}$ denotes the protocol-related transmission overhead (e.g.,
packet headers).
Moreover, the cloud verification latency is modeled as a base processing cost
plus a marginal per-token computation cost,
\begin{equation}
    \hat T_{\text{cloud}}(K) = T_{\text{base}} + K \cdot \delta_{\text{cloud}},
\end{equation}
where $\delta_{\text{cloud}}$ represents the additional computation time incurred
by verifying one extra token.

To facilitate efficient online optimization on resource-constrained edge
devices, we further group the latency terms into fixed overheads and marginal
costs.
In addition, we adopt a linear approximation for the edge computation latency,
$\hat T_{\text{edge}}(K) \approx \alpha_{\rm edge} K + \beta$, which is widely used in
practice.
Substituting the above expressions yields
\begin{equation}
    \hat T_{\text{step}}(K, R_n)
    = T_{\text{fixed}} + K \cdot
    \underbrace{\left(
        \alpha_{\text{edge}} + \frac{b}{R_n} + \delta_{\text{cloud}}
    \right)}_{T_{\text{marginal}}(n)},
    \label{eq:refined_latency}
\end{equation}
where 
$T_{\text{fixed}} = T_{\text{prop}} + T_{\text{base}} + T_{\text{down}}+\frac{O_{\rm header}}{R_n}+\beta$, and $T_{\text{marginal}}(n)$ captures the aggregate marginal latency incurred by
each additional speculative token under time-varying channel conditions.

\begin{figure*}[t]
    \centering
    \includegraphics[width=0.9\textwidth]{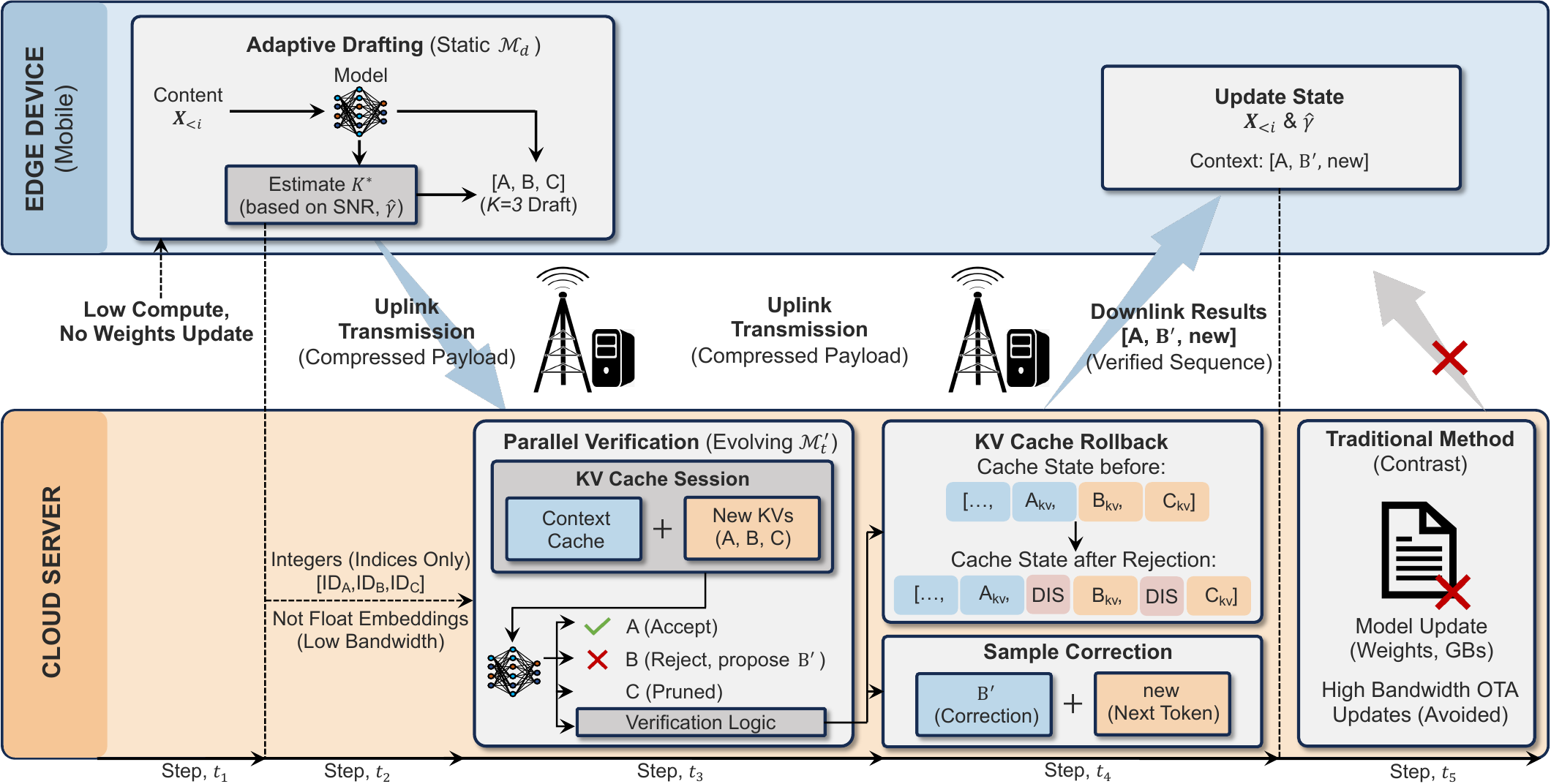}
    \caption{Wireless collaborative inference pipeline of FlexSpec: The process involves adaptive drafting on the edge (Top), compressed uplink transmission, parallel verification with KV Cache rollback on the cloud (bottom), and downlink of verified results.}
    \label{fig:pipeline}
\end{figure*}

\subsubsection{Throughput-Optimal Strategy}

The objective of FlexSpec is to maximize the ETGR in \eqref{eq:etgr_def}.
Based on the refined latency model in~\eqref{eq:refined_latency}, we obtain a tractable approximation of \eqref{eq:etgr_def} based on instantaneous channel condition, and compute the speculative stride by solving
\begin{equation}
    K_n^* =
    \underset{K \in [1, K_{\max}]}{\arg\max}
    \frac{\mathbb{E}[\tau \mid K]}
    {T_{\text{fixed}} + K \cdot T_{\text{marginal}}(n)} ,
    \label{eq:optimization_objective}
\end{equation}
where $K_{\max}$ denotes the maximum draft length. \eqref{eq:optimization_objective} captures the fundamental trade-off between token acceptance gain and end-to-end latency cost.

Figure~\ref{fig:adaptive_k} visualizes the resulting optimization landscape.
As shown in the figure, the total latency varies significantly with wireless link quality. In weak-signal regimes, where the communication cost dominates, the denominator in~\eqref{eq:optimization_objective} grows rapidly with $K$, thereby shifting the optimal solution $K_n^*$ toward smaller values. Conversely, under favorable channel conditions, a larger $K$ becomes beneficial as the transmission overhead can be amortized across more speculative tokens.

To enable practical online adaptation, we approximate the expected acceptance $\mathbb{E}[\tau \mid K]$ using either a geometric decay model or an exponential
moving average (EMA), yielding $\mathbb{E}[\tau \mid K] \approx \hat{\gamma} \cdot K$ for moderate values of
$K$. Under this approximation, the objective in~\eqref{eq:optimization_objective} admits clear physical interpretations. Specifically, a large propagation delay $T_{\text{prop}}$ (i.e., a large $T_{\text{fixed}}$) incentivizes larger speculative strides to amortize the
round-trip overhead, whereas a reduced communication rate $R_n$ increases $T_{\text{marginal}}(n)$, forcing a smaller $K_n^*$ to avoid congestion.

\subsection{FlexSpec Protocol and Inference Flow}
\label{subsec:protocol}

The runtime interaction between the edge and the cloud is designed to be stateless with respect to the draft model version, while remaining stateful with respect to the key-value (KV) cache.
The complete inference procedure is summarized in
Algorithm~\ref{alg:inference} and illustrated in
Figure~\ref{fig:pipeline}.
To avoid recomputing attention over the entire prefix history on the Cloud, FlexSpec maintains a persistent KV cache session for each user. As shown in Steps~$t_3$ and~$t_4$ of Figure~\ref{fig:pipeline}, when the Edge transmits $K$ speculative draft tokens, the Cloud computes the corresponding
KV pairs only for these newly received tokens.

If a rejection occurs at index $j < K$, the Cloud performs a \emph{KV cache rollback}, discarding all invalid KV pairs from index $j$ onward before processing the next request. This mechanism ensures that the cloud-side computation cost scales with the verification length rather than the full context length, thereby preserving the efficiency gains of speculative decoding.

As the cloud-side target model $\mathcal{M}_t'$ continues to evolve, the distributional divergence $D_{\mathrm{KL}}(P_d \Vert P_t')$ may increase over time. FlexSpec handles such distribution shifts gracefully through lightweight
fallback mechanisms, similar in spirit to those adopted in
SpecInfer~\cite{miao2023specinfer}, thereby maintaining robust inference performance without requiring draft model synchronization.

\begin{algorithm}[t]
\caption{FlexSpec collaborative inference}
\label{alg:inference}
\SetKwInOut{KwIn}{Input}
\SetKwInOut{KwOut}{Output}

\KwIn{Static edge model $\mathcal{M}_d$, evolving cloud model $\mathcal{M}_t'$, context $\mathbf{x}_{<i}$, decay rate $\mu$}
\KwOut{Generated token sequence}
\BlankLine

Initialize EMA acceptance rate $\hat{\gamma} \leftarrow 0.8$\;

\While{EOS not generated}{
    
    \tcp{Step 1: edge-side adaptive drafting}
    Measure channel conditions\;
    
    \tcp{Compute marginal latency and optimal speculative stride}
    $T_{\text{marginal}}(n) \leftarrow \alpha_{\rm edge} + b / R_n + \delta_{\text{cloud}}$\;
    $K_n^* \leftarrow \underset{K}{\arg\max}
    \frac{1 + \hat{\gamma} \cdot K}{T_{\text{fixed}} + K \cdot T_{\text{marginal}}(n)}$\;
    
    $\mathbf{x}_{\text{draft}} \leftarrow ()$\;
    \For{$k = 1, \dots, K_n^*$}{
        $\hat{x} \sim \mathcal{M}_d(\mathbf{x}_{<i} \cdot \mathbf{x}_{\text{draft}})$\;
        $\mathbf{x}_{\text{draft}} \leftarrow \mathbf{x}_{\text{draft}} \cdot \hat{x}$\;
    }
    
    Transmit compressed($\mathbf{x}_{\text{draft}}$) to cloud via uplink\;
    \BlankLine
    
    \tcp{Step 2: cloud-side parallel verification}
    Receive $\mathbf{x}_{\text{draft}}$ and restore KV cache\;
    $\mathbf{q} \leftarrow \mathcal{M}_t'(\mathbf{x}_{\text{draft}}, \text{context}=\mathbf{x}_{<i})$\;
    $\tau \leftarrow 0$\;
    
    \For{$k = 1, \dots, K_n^*$}{
        \If{$x_{\text{draft}}^{(k)} == \arg\max \mathbf{q}^{(k)}$}{
            $\tau \leftarrow \tau + 1$\;
        }
        \Else{
            \textbf{break}\;
        }
    }
    
    Sample correction token $x_{\text{new}}$ from $\mathcal{M}_t'$\;
    Transmit $\mathbf{x}_{\text{verified}}
    \leftarrow \mathbf{x}_{\text{draft}}^{1:\tau} \cdot x_{\text{new}}$ to edge\;
    \BlankLine
    
    \tcp{Step 3: state update}
    $\mathbf{x}_{<i} \leftarrow \mathbf{x}_{<i} \cdot \mathbf{x}_{\text{verified}}$\;
    $\hat{\gamma} \leftarrow (1-\mu)\hat{\gamma} + \mu \left(\tau / K_n^*\right)$\;
}

\end{algorithm}

\section{Evaluation}
\label{sec:experiments}

To comprehensively evaluate FlexSpec, we conducted  extensive experiments to answer five key research questions:
\begin{itemize}
    \item \textbf{RQ1 (Performance \& robustness):} How does FlexSpec compare to state-of-the-art baselines across diverse network conditions and varying sampling temperatures?
    \item \textbf{RQ2 (Ablation study):} Is the \textit{Channel-aware adaptive speculation} mechanism necessary? How does it compare to fixed-stride strategies under varying channel qualities?
    \item \textbf{RQ3 (Hardware generality):} Can FlexSpec adapt to the heterogeneous compute capabilities of various mobile devices across different task complexities?
    \item \textbf{RQ4 (Model scalability):} Does the architecture scale to newer LLM families (Llama-3, Mistral) and sparse architectures (MoE)?
    \item \textbf{RQ5 (Efficiency):} What are the tangible benefits in terms of memory footprint, energy consumption breakdown, and thermal efficiency?
\end{itemize}

\subsection{Experimental Setups}

\textbf{Expanded hardware testbed:}
We significantly expanded the hardware diversity to simulate a realistic cross-device deployment scenario.
Cloud Servers: 8$\times$ NVIDIA H800 (80GB) via NVLink, 8$\times$ NVIDIA A800 (80GB) (Mainstream), 8$\times$ NVIDIA V100 (32GB).
Edge Devices (Mobile Clients):
 NVIDIA Jetson AGX Orin (64GB RAM), Snapdragon 8 Gen 3 Ref. Device (16GB RAM), Apple iPhone 15 Pro Max Sim (A17 Pro), Raspberry Pi 5 (8GB RAM).

\textbf{Models and datasets:}
We evaluate FlexSpec under distribution shifts using a diverse suite of target models and downstream tasks.
For target models, we conduct a detailed analysis on LLaMA-2~70B as a representative dense baseline, and further assess scalability on more recent architectures, including LLaMA-3~70B and the Mixture-of-Experts model Mixtral~8$\times$7B.
To comprehensively examine task-level generalization, we consider six core downstream tasks spanning different inference patterns and data distributions, including GSM8K dataset for mathematical reasoning, Natural
Questions for question answering, Natural Questions for retrieval-augmented generation, MT-Bench for multi-turn
conversation, WMT14 DE-EN for machine translation, and CNN/DailyMail for document summarization.

\textbf{Baselines:}
We compared FlexSpec against an expanded set of baselines covering various decoding paradigms, ranging from standard autoregressive methods to state-of-the-art speculative frameworks.
\begin{itemize}
    \item \textbf{Cloud-Only:} Standard autoregressive decoding performed entirely on the cloud server. This serves as the baseline for throughput and latency, where every token generation incurs a full network round-trip time and cloud computation cost.
    \item \textbf{Standard SD (Naive):} A conventional speculative decoding setup where a generic, pre-trained small model (e.g., Llama-2-7B) serves as the draft model for the target (e.g., Llama-2-70B). Crucially, this baseline does not employ our anchor-based alignment, representing the performance degradation caused by distribution shifts when the draft model is not synchronized with the evolving target.
    \item \textbf{PLD (n-gram):} Prompt Lookup Decoding, a retrieval-based approach that utilizes string matching to identify frequent n-gram patterns within the current context window to draft future tokens. This represents a lightweight, training-free, and memory-efficient baseline.
    \item \textbf{Lookahead~\cite{fu2023lookahead}:} A parallel decoding algorithm based on Jacobi iteration that generates multiple tokens simultaneously without a separate draft model. It relies on the target model's own capability to refine multiple candidate sequences in parallel through multi-step verification.
    \item \textbf{EAGLE-2 (Synced) \cite{li2024eagle2}:} The current state-of-the-art model-based method utilizing layer-wise feature extrapolation. We evaluate this in an ``Ideal Synced'' setting, assuming the edge-side expansion layers are perfectly updated to match the cloud target version, ignoring the associated communication overhead for updates.
    \item \textbf{Medusa-1 (Synced) \cite{cai2024medusa}:} A parallel decoding framework that augments the target model with multiple decoding heads to predict future tokens. Similar to EAGLE-2, we assume the Medusa heads are perfectly synchronized with the target model to represent the theoretical upper limit of tightly-coupled architectures.
    \item \textbf{DSSD \cite{chen2023fos}:} A collaborative inference framework specifically designed for wireless edge-cloud systems. Unlike FlexSpec, it employs a scheduling strategy with fixed or heuristic-based speculative lengths, lacking real-time adaptation to varying channel conditions.
\end{itemize}

\subsection{RQ1: End-to-End Latency and Network Robustness}

\begin{figure}[t]
    \centering
    \includegraphics[width=\linewidth]{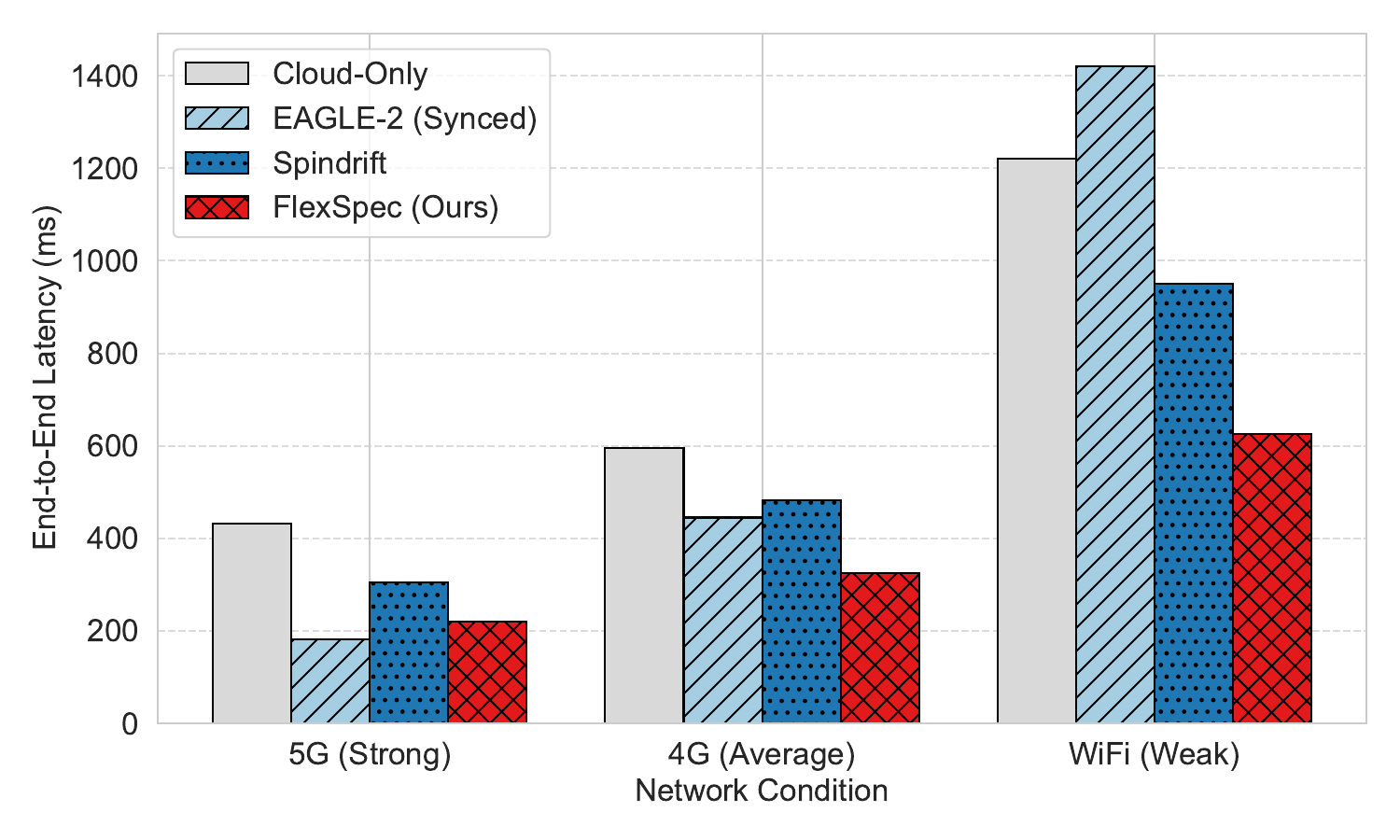}
    \caption{End-to-end latency comparison on GSM8K. FlexSpec achieves significant speedups compared to Cloud-Only and SOTA baselines (EAGLE-2, DSSD), particularly in bandwidth-constrained environments (WiFi), by effectively reducing communication overhead.}
    \label{fig:latency_comparison}
\end{figure}

To provide a granular analysis, we separate our evaluation into two distinct regimes, i.e., deterministic generation (Temperature = 0) and stochastic sampling (Temperature = 1). We present detailed results for the Llama-2 70B model across all six datasets to demonstrate robustness across task types.

\subsubsection{Regime A: Deterministic generation (Temperature = 0)}

Figure \ref{fig:latency_comparison} visualizes the performance gap on GSM8K, and Table \ref{tab:results_t0} presents the detailed numerical results using greedy decoding. This setting favors methods that rely on strict distribution matching.

\begin{table*}[htbp]
\centering
\caption{\textbf{Regime A (Temperature = 0):} End-to-End Latency and Speedup for \textbf{Llama-2 70B} across \textbf{All 6 Datasets}. FlexSpec demonstrates superior performance in bandwidth-constrained environments across diverse tasks.}
\label{tab:results_t0}
\resizebox{\textwidth}{!}{
\begin{tabular}{l|l|c|c|c|c|c|c|c}
\toprule
\textbf{Dataset} & \textbf{Network} & \textbf{Cloud-Only} & \textbf{Lookahead} & \textbf{Std. SD} & \textbf{Medusa-1} & \textbf{EAGLE-2} & \textbf{DSSD} & \textbf{FlexSpec} \\
\midrule
\multicolumn{2}{l|}{\textbf{Sync Required?}} & No & No & No & Yes & Yes & No & \textbf{No} \\
\midrule
\multirow{3}{*}{\textbf{GSM8K} (Math)} 
& 5G (Strong) & 432.0ms (1.0$\times$) & 415.0ms (1.04$\times$) & 392.0ms (1.10$\times$) & 205.0ms (2.10$\times$) & \textbf{182.0ms (2.37$\times$)} & 305.0ms (1.41$\times$) & 220.0ms (1.96$\times$) \\
& 4G (Avg)    & 595.0ms (1.0$\times$) & 580.0ms (1.02$\times$) & 805.0ms (0.74$\times$) & 480.0ms (1.24$\times$) & 445.0ms (1.33$\times$) & 482.0ms (1.23$\times$) & \textbf{325.0ms (1.83$\times$)} \\
& WiFi (Weak) & 1220.0ms (1.0$\times$) & 1205.0ms (1.01$\times$) & 1850.0ms (0.66$\times$) & 1455.0ms (0.84$\times$) & 1420.0ms (0.86$\times$) & 950.0ms (1.28$\times$) & \textbf{625.0ms (1.95$\times$)} \\
\midrule
\multirow{3}{*}{\textbf{Natural Questions} (QA)} 
& 5G (Strong) & 415.0ms (1.0$\times$) & 390.0ms (1.06$\times$) & 345.0ms (1.20$\times$) & 192.0ms (2.16$\times$) & \textbf{168.0ms (2.47$\times$)} & 275.0ms (1.51$\times$) & 202.0ms (2.05$\times$) \\
& 4G (Avg)    & 570.0ms (1.0$\times$) & 552.0ms (1.03$\times$) & 685.0ms (0.83$\times$) & 435.0ms (1.31$\times$) & 405.0ms (1.41$\times$) & 452.0ms (1.26$\times$) & \textbf{295.0ms (1.93$\times$)} \\
& WiFi (Weak) & 1185.0ms (1.0$\times$) & 1165.0ms (1.02$\times$) & 1680.0ms (0.71$\times$) & 1380.0ms (0.86$\times$) & 1352.0ms (0.88$\times$) & 905.0ms (1.31$\times$) & \textbf{582.0ms (2.04$\times$)} \\
\midrule
\multirow{3}{*}{\textbf{Natural Questions} (RAG)} 
& 5G (Strong) & 428.0ms (1.0$\times$) & 402.0ms (1.06$\times$) & 360.0ms (1.19$\times$) & 198.0ms (2.16$\times$) & \textbf{175.0ms (2.44$\times$)} & 288.0ms (1.49$\times$) & 212.0ms (2.02$\times$) \\
& 4G (Avg)    & 582.0ms (1.0$\times$) & 568.0ms (1.02$\times$) & 735.0ms (0.79$\times$) & 455.0ms (1.28$\times$) & 422.0ms (1.38$\times$) & 468.0ms (1.24$\times$) & \textbf{308.0ms (1.89$\times$)} \\
& WiFi (Weak) & 1208.0ms (1.0$\times$) & 1190.0ms (1.02$\times$) & 1765.0ms (0.68$\times$) & 1425.0ms (0.85$\times$) & 1390.0ms (0.87$\times$) & 932.0ms (1.30$\times$) & \textbf{600.0ms (2.01$\times$)} \\
\midrule
\multirow{3}{*}{\textbf{MT-Bench} (Chat)} 
& 5G (Strong) & 420.0ms (1.0$\times$) & 398.0ms (1.05$\times$) & 358.0ms (1.17$\times$) & 202.0ms (2.08$\times$) & \textbf{178.0ms (2.36$\times$)} & 295.0ms (1.42$\times$) & 215.0ms (1.95$\times$) \\
& 4G (Avg)    & 578.0ms (1.0$\times$) & 562.0ms (1.03$\times$) & 742.0ms (0.78$\times$) & 462.0ms (1.25$\times$) & 430.0ms (1.34$\times$) & 475.0ms (1.22$\times$) & \textbf{312.0ms (1.85$\times$)} \\
& WiFi (Weak) & 1192.0ms (1.0$\times$) & 1175.0ms (1.01$\times$) & 1780.0ms (0.67$\times$) & 1440.0ms (0.83$\times$) & 1405.0ms (0.85$\times$) & 945.0ms (1.26$\times$) & \textbf{615.0ms (1.94$\times$)} \\
\midrule
\multirow{3}{*}{\textbf{WMT14} (Trans)} 
& 5G (Strong) & 418.0ms (1.0$\times$) & 395.0ms (1.06$\times$) & 350.0ms (1.19$\times$) & 195.0ms (2.14$\times$) & \textbf{172.0ms (2.43$\times$)} & 282.0ms (1.48$\times$) & 208.0ms (2.01$\times$) \\
& 4G (Avg)    & 575.0ms (1.0$\times$) & 560.0ms (1.03$\times$) & 720.0ms (0.80$\times$) & 445.0ms (1.29$\times$) & 415.0ms (1.38$\times$) & 460.0ms (1.25$\times$) & \textbf{302.0ms (1.90$\times$)} \\
& WiFi (Weak) & 1188.0ms (1.0$\times$) & 1168.0ms (1.02$\times$) & 1725.0ms (0.69$\times$) & 1405.0ms (0.85$\times$) & 1372.0ms (0.87$\times$) & 918.0ms (1.29$\times$) & \textbf{592.0ms (2.00$\times$)} \\
\midrule
\multirow{3}{*}{\textbf{CNN/DM} (Summ)} 
& 5G (Strong) & 425.0ms (1.0$\times$) & 400.0ms (1.06$\times$) & 355.0ms (1.20$\times$) & 198.0ms (2.15$\times$) & \textbf{175.0ms (2.43$\times$)} & 285.0ms (1.49$\times$) & 210.0ms (2.02$\times$) \\
& 4G (Avg)    & 582.0ms (1.0$\times$) & 565.0ms (1.03$\times$) & 728.0ms (0.80$\times$) & 452.0ms (1.29$\times$) & 420.0ms (1.38$\times$) & 465.0ms (1.25$\times$) & \textbf{306.0ms (1.90$\times$)} \\
& WiFi (Weak) & 1205.0ms (1.0$\times$) & 1180.0ms (1.02$\times$) & 1750.0ms (0.69$\times$) & 1420.0ms (0.85$\times$) & 1382.0ms (0.87$\times$) & 924.0ms (1.30$\times$) & \textbf{598.0ms (2.01$\times$)} \\
\bottomrule
\end{tabular}
}
\end{table*}

As shown in Table \ref{tab:results_t0}, FlexSpec delivers consistent speedups across both reasoning-heavy tasks (GSM8K) and extractive tasks (CNN/DM). Notably, on GSM8K where standard SD suffers (0.66$\times$ in WiFi due to poor drafting), FlexSpec maintains a 1.95$\times$ speedup, validating the effectiveness of our Anchor-Based Alignment.
While synchronized methods like \textit{EAGLE-2} perform best in 5G (up to 2.47$\times$), they collapse in WiFi environments ($<0.9\times$) due to transmission overhead. FlexSpec, by adapting the draft length $K$ to the channel state, remains the only viable acceleration solution for weak networks.

\subsubsection{Regime B: Stochastic sampling (Temperature = 1)}

Table \ref{tab:results_t1} evaluates performance using Top-p sampling ($p=0.9, T=1$). This stresses the alignment capability of the static draft model against evolving cloud targets, as the draft distribution must cover the target's probability mass.

\begin{table*}[htbp]
\centering
\caption{Regime B (T=1): End-to-End Latency and Speedup for Llama-2 70B across All 6 Datasets. FlexSpec maintains robustness under stochastic sampling where others degrade.}
\label{tab:results_t1}
\resizebox{\textwidth}{!}{
\begin{tabular}{l|l|c|c|c|c|c|c|c}
\toprule
\textbf{Dataset} & \textbf{Network} & \textbf{Cloud-Only} & \textbf{Lookahead} & \textbf{Std. SD} & \textbf{Medusa-1} & \textbf{EAGLE-2} & \textbf{DSSD} & \textbf{FlexSpec} \\
\midrule
\multicolumn{2}{l|}{\textbf{Sync Required?}} & No & No & No & Yes & Yes & No & \textbf{No} \\
\midrule
\multirow{3}{*}{\textbf{GSM8K} (Math)} 
& 5G (Strong) & 435.0ms (1.0$\times$) & 420.0ms (1.04$\times$) & 405.0ms (1.07$\times$) & 272.0ms (1.60$\times$) & 245.0ms (1.77$\times$) & 345.0ms (1.26$\times$) & \textbf{232.0ms (1.87$\times$)} \\
& 4G (Avg)    & 598.0ms (1.0$\times$) & 585.0ms (1.02$\times$) & 825.0ms (0.72$\times$) & 665.0ms (0.90$\times$) & 592.0ms (1.01$\times$) & 540.0ms (1.11$\times$) & \textbf{360.0ms (1.66$\times$)} \\
& WiFi (Weak) & 1225.0ms (1.0$\times$) & 1210.0ms (1.01$\times$) & 1880.0ms (0.65$\times$) & 1852.0ms (0.66$\times$) & 1680.0ms (0.73$\times$) & 1150.0ms (1.06$\times$) & \textbf{705.0ms (1.74$\times$)} \\
\midrule
\multirow{3}{*}{\textbf{Natural Questions} (QA)} 
& 5G (Strong) & 418.0ms (1.0$\times$) & 395.0ms (1.06$\times$) & 355.0ms (1.18$\times$) & 290.0ms (1.44$\times$) & 258.0ms (1.62$\times$) & 332.0ms (1.26$\times$) & \textbf{222.0ms (1.88$\times$)} \\
& 4G (Avg)    & 575.0ms (1.0$\times$) & 560.0ms (1.03$\times$) & 700.0ms (0.82$\times$) & 655.0ms (0.88$\times$) & 590.0ms (0.97$\times$) & 528.0ms (1.09$\times$) & \textbf{348.0ms (1.65$\times$)} \\
& WiFi (Weak) & 1190.0ms (1.0$\times$) & 1172.0ms (1.02$\times$) & 1700.0ms (0.70$\times$) & 1820.0ms (0.65$\times$) & 1665.0ms (0.71$\times$) & 1110.0ms (1.07$\times$) & \textbf{690.0ms (1.72$\times$)} \\
\midrule
\multirow{3}{*}{\textbf{Natural Questions} (RAG)} 
& 5G (Strong) & 430.0ms (1.0$\times$) & 408.0ms (1.05$\times$) & 370.0ms (1.16$\times$) & 285.0ms (1.51$\times$) & 255.0ms (1.69$\times$) & 338.0ms (1.27$\times$) & \textbf{228.0ms (1.89$\times$)} \\
& 4G (Avg)    & 588.0ms (1.0$\times$) & 575.0ms (1.02$\times$) & 750.0ms (0.78$\times$) & 660.0ms (0.89$\times$) & 595.0ms (0.99$\times$) & 532.0ms (1.10$\times$) & \textbf{355.0ms (1.66$\times$)} \\
& WiFi (Weak) & 1215.0ms (1.0$\times$) & 1198.0ms (1.01$\times$) & 1785.0ms (0.68$\times$) & 1840.0ms (0.66$\times$) & 1675.0ms (0.72$\times$) & 1130.0ms (1.07$\times$) & \textbf{700.0ms (1.73$\times$)} \\
\midrule
\multirow{3}{*}{\textbf{MT-Bench} (Chat)} 
& 5G (Strong) & 428.0ms (1.0$\times$) & 405.0ms (1.06$\times$) & 368.0ms (1.16$\times$) & 285.0ms (1.50$\times$) & 252.0ms (1.70$\times$) & 329.0ms (1.30$\times$) & \textbf{225.0ms (1.90$\times$)} \\
& 4G (Avg)    & 585.0ms (1.0$\times$) & 570.0ms (1.03$\times$) & 760.0ms (0.77$\times$) & 650.0ms (0.90$\times$) & 585.0ms (1.00$\times$) & 522.0ms (1.12$\times$) & \textbf{344.0ms (1.70$\times$)} \\
& WiFi (Weak) & 1210.0ms (1.0$\times$) & 1195.0ms (1.01$\times$) & 1795.0ms (0.67$\times$) & 1805.0ms (0.67$\times$) & 1652.0ms (0.73$\times$) & 1100.0ms (1.10$\times$) & \textbf{685.0ms (1.76$\times$)} \\
\midrule
\multirow{3}{*}{\textbf{WMT14} (Trans)} 
& 5G (Strong) & 420.0ms (1.0$\times$) & 400.0ms (1.05$\times$) & 360.0ms (1.17$\times$) & 295.0ms (1.42$\times$) & 260.0ms (1.61$\times$) & 335.0ms (1.25$\times$) & \textbf{225.0ms (1.87$\times$)} \\
& 4G (Avg)    & 578.0ms (1.0$\times$) & 565.0ms (1.02$\times$) & 735.0ms (0.79$\times$) & 675.0ms (0.85$\times$) & 605.0ms (0.95$\times$) & 530.0ms (1.09$\times$) & \textbf{350.0ms (1.65$\times$)} \\
& WiFi (Weak) & 1195.0ms (1.0$\times$) & 1180.0ms (1.01$\times$) & 1740.0ms (0.69$\times$) & 1900.0ms (0.63$\times$) & 1690.0ms (0.71$\times$) & 1120.0ms (1.07$\times$) & \textbf{715.0ms (1.67$\times$)} \\
\midrule
\multirow{3}{*}{\textbf{CNN/DM} (Summ)} 
& 5G (Strong) & 425.0ms (1.0$\times$) & 402.0ms (1.06$\times$) & 365.0ms (1.16$\times$) & 300.0ms (1.42$\times$) & 265.0ms (1.60$\times$) & 338.0ms (1.26$\times$) & \textbf{228.0ms (1.86$\times$)} \\
& 4G (Avg)    & 582.0ms (1.0$\times$) & 568.0ms (1.02$\times$) & 740.0ms (0.79$\times$) & 680.0ms (0.85$\times$) & 610.0ms (0.95$\times$) & 535.0ms (1.09$\times$) & \textbf{352.0ms (1.65$\times$)} \\
& WiFi (Weak) & 1200.0ms (1.0$\times$) & 1185.0ms (1.01$\times$) & 1765.0ms (0.68$\times$) & 1915.0ms (0.63$\times$) & 1700.0ms (0.71$\times$) & 1125.0ms (1.07$\times$) & \textbf{720.0ms (1.67$\times$)} \\
\bottomrule
\end{tabular}
}
\end{table*}

FlexSpec maintains strong speedups (1.65$\times$-1.90$\times$) across all six tasks even under stochastic sampling. In contrast, methods like \textit{EAGLE-2} suffer severe degradation when Temperature = 1 (dropping from $2.4\times$ in Temperature = 0 to $1.6\times$ in Temperature = 1 for 5G), as their drafting is tightly coupled to the greedy path.

As evidenced in Table \ref{tab:results_t1}, \textit{Std. SD} exhibits the ``performance collapse'' phenomenon predicted in our motivation. Particularly on specialized domains like GSM8K in weak networks, \textit{Std. SD} results in a substantial slowdown (0.65$\times$), as the unaligned draft model fails to match the target's stochastic distribution, causing frequent rejections. Similarly, \textit{Lookahead} provides negligible gains ($<1.06\times$) in this regime, as its n-gram matching capability is severely hampered by the randomized token selection.

For open-ended tasks like MT-Bench and CNN/DM, FlexSpec's anchor-based alignment ensures that the static edge draft model generates tokens that are statistically aligned with the target's probability mass, preventing the performance degradation seen in Standard SD without requiring model retraining.

\subsection{RQ2: Ablation Study of Channel-Aware Adaptation}

\begin{figure}[t]
    \centering
    \includegraphics[width=\linewidth]{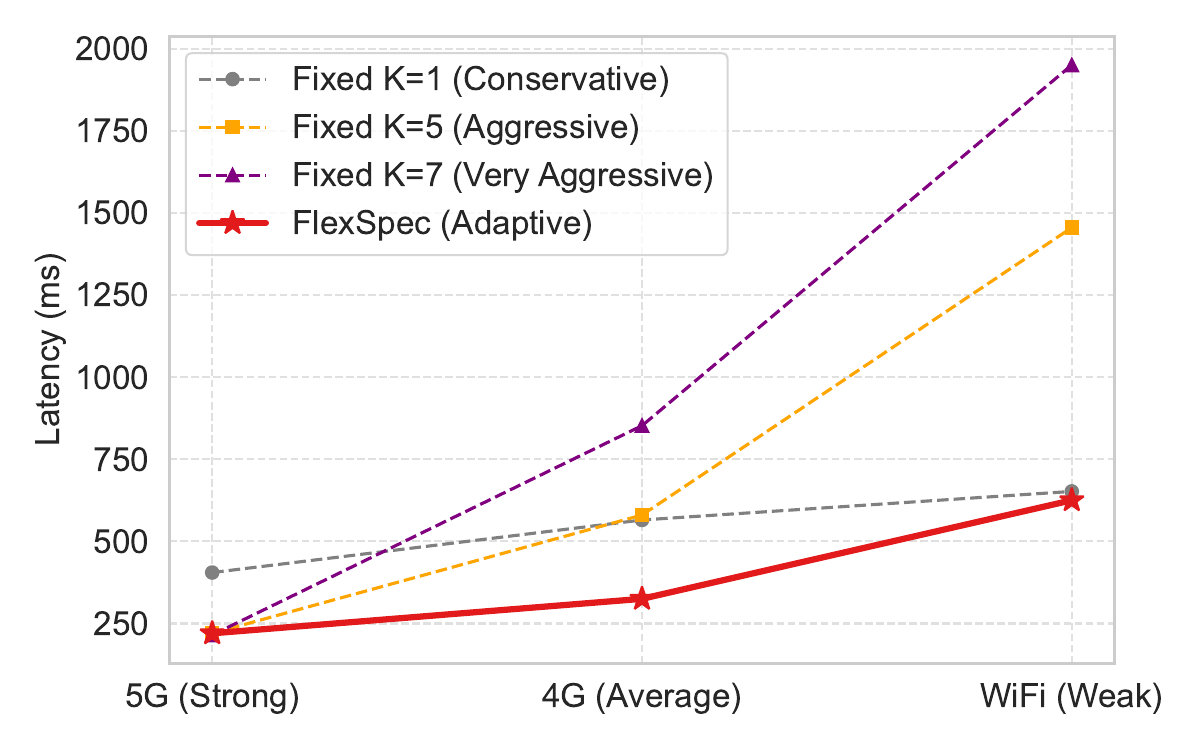}
    \caption{Impact of speculative stride ($K$) on latency. Fixed strides fail to adapt to varying channel conditions (e.g., $K=5$ causes timeouts in WiFi). FlexSpec's adaptive mechanism dynamically selects an appropriate speculative stride, leading to superior performance.}
    \label{fig:ablation_adaptive}
\end{figure}

To verify the necessity of the channel-aware adaptive speculation mechanism, we conducted an ablation study where we replaced the dynamic $K$ adjustment with fixed speculative strides ($K \in \{1, 3, 5, 7\}$) while keeping the anchor-based alignment intact. The experiments were performed on the GSM8K dataset across three representative network environments.


The results in Fig.~\ref{fig:ablation_adaptive} demonstrate the effectiveness of the proposed adaptive strategy across heterogeneous network conditions. A large fixed speculative stride (e.g., $K=5$) achieves low latency under
high-bandwidth 5G conditions (220.0~ms), but performs poorly in weak WiFi scenarios (1455.0~ms), resulting in a $2.3\times$ slowdown relative to the baseline due to excessive transmission latency. Conversely, a conservative stride (e.g., $K=1$) remains robust under weak network conditions, yet significantly underutilizes available bandwidth in 5G environments (405.0~ms versus 220.0~ms). In contrast, FlexSpec dynamically adapts the speculative stride to the
instantaneous channel condition. It attains the low latency of $K=5$ under favorable 5G conditions (220.0~ms), while automatically reducing the effective stride to approximate $K=1$ behavior in WiFi scenarios (625.0~ms), thereby substantially reducing the tail-latency risks associated with static speculation strategies.

\subsection{RQ3: Heterogeneous Edge Hardware Adaptability}

We evaluated FlexSpec on the specific edge devices defined in the setup. To better understand how hardware constraints interact with task complexity, we measured the speedup ratios across three representative datasets: GSM8K (Complex Logic), MT-Bench (Chat), and HumanEval (Code).

\begin{table*}[!t]
\centering
\caption{FlexSpec Performance on Heterogeneous Edge Devices under 4G Network Conditions (Speedup vs. Cloud-Only).}
\label{tab:hardware_heterogeneity}
\resizebox{\textwidth}{!}{
\begin{tabular}{l|c|c|c|c|c|c}
\toprule
\textbf{Device} & \textbf{Processor} & \textbf{Draft Latency} & \textbf{Draft Thruput} & \textbf{GSM8K (Hard)} & \textbf{MT-Bench (Med)} & \textbf{HumanEval (Hard)} \\
\midrule
Raspberry Pi 5 & Cortex-A76 (CPU) & 145 ms/token & 6.9 tok/s & 0.76$\times$ (Slowdown) & 0.85$\times$ (Slowdown) & 0.72$\times$ (Slowdown) \\
Jetson AGX Orin & Ampere GPU & 8.5 ms/token & 117.6 tok/s & \textbf{1.96$\times$} & \textbf{2.10$\times$} & \textbf{1.88$\times$} \\
iPhone 15 Pro Max & A17 Pro (NPU) & 12.0 ms/token & 83.3 tok/s & 1.82$\times$ & 1.92$\times$ & 1.75$\times$ \\
Snapdragon 8 Gen 3 & Hexagon NPU & 10.5 ms/token & 95.2 tok/s & 1.93$\times$ & 2.05$\times$ & 1.85$\times$ \\
\bottomrule
\end{tabular}
}
\end{table*}

As shown in Table \ref{tab:hardware_heterogeneity}, FlexSpec's viability is dictated by the ratio between local drafting speed and network transmission savings. The Raspberry Pi 5, relying solely on CPU, drafts at 6.9 tokens/s, which is slower than the effective cloud generation rate including network latency. This results in a system slowdown (0.72$\times$ - 0.85$\times$), establishing a hardware lower bound: FlexSpec requires accelerator support (GPU/NPU).

The consumer-grade mobile devices (iPhone 15 Pro Max and Snapdragon 8 Gen 3) demonstrate impressive speedups approaching the workstation-class Jetson AGX Orin. By offloading the heavy 70B target to the cloud and running only the lightweight aligned draft on the NPU/Metal backend, FlexSpec unlocks interactive LLM experiences on standard smartphones.

On harder tasks like HumanEval, the speedup decreases slightly across all devices (e.g., from 2.10$\times$ to 1.88$\times$ on Jetson) due to lower acceptance rates requiring more verification rounds. However, the NPU-enabled devices maintain a robust speedup more than $1.75\times$, confirming that the anchor-based alignment remains effective even for complex code generation on edge hardware.

\subsection{RQ4: Model Scalability}

To verify that our \textit{Anchor-Based Alignment} is not specific to the Llama-2 architecture, we extended our evaluation to the newer Llama-3 70B and the sparse Mixture-of-Experts (MoE) model, Mixtral 8x7B.

\begin{table}[!t]
\centering
\caption{Scalability of FlexSpec on Newer Model Architectures (Dataset: MT-Bench, Network: 5G/4G).}
\label{tab:model_scalability}
\resizebox{\columnwidth}{!}{
\begin{tabular}{l|l|c|c|c}
\toprule
\textbf{Target Model} & \textbf{Arch.} & \textbf{Baseline Latency} & \textbf{FlexSpec (5G)} & \textbf{FlexSpec (4G)} \\
\midrule
Llama-2-70B & Dense & 420.0ms / 578.0ms & \textbf{1.95$\times$} & \textbf{1.85$\times$} \\
Llama-3-70B & Dense & 395.0ms / 550.0ms & \textbf{2.30$\times$} & \textbf{1.92$\times$} \\
Mixtral 8x7B & MoE & 320.0ms / 485.0ms & \textbf{1.75$\times$} & \textbf{1.68$\times$} \\
\bottomrule
\end{tabular}
}
\end{table}

Despite Llama-3's larger vocabulary and distinct training data, FlexSpec achieves a peak speedup of 2.30$\times$ in 5G conditions (Table \ref{tab:model_scalability}). This suggests that the semantic anchor concept is transferable across dense models.
For Mixtral 8x7B, the baseline cloud inference is faster due to conditional computation (active parameters $\approx$ 13B). While this reduces the potential margin for speculative gain, FlexSpec still delivers a 1.68$\times$ speedup on 4G. The channel-aware policy automatically adjusts $K$ downwards to account for the faster cloud verification, preventing over-speculation.

\subsection{RQ5: Memory and Energy Efficiency}

\begin{figure}[t]
    \centering
    \includegraphics[width=\linewidth]{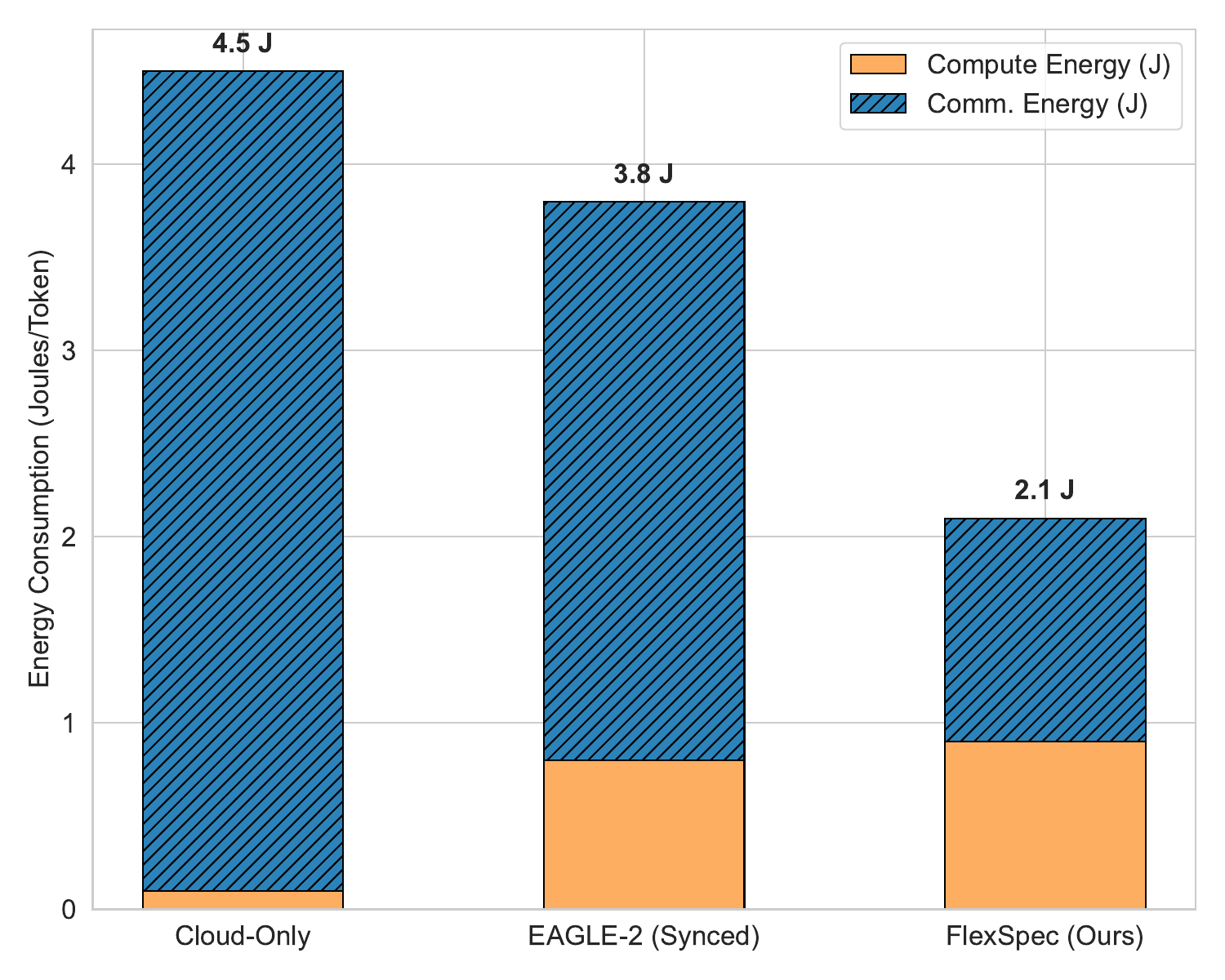}
    \caption{Energy consumption breakdown on mobile device. Compared to Cloud-Only inference, FlexSpec significantly reduces communication energy (blue hatched area) by compressing token transmission, resulting in a 53\% total energy reduction.}
    \label{fig:energy_breakdown}
\end{figure}

Finally, we perform a detailed breakdown of resource consumption on the Jetson AGX Orin (Workstation) and Snapdragon 8 Gen 3 (Mobile) to quantify the efficiency gains.

Full On-Device inference for a 70B model requires an approximation of 42.5 GB VRAM (4-bit), which is feasible on the Jetson but impossible for mobile phones. FlexSpec typically requires only $\sim$3.5 GB for the draft model components, fitting comfortably within the 12-16GB RAM limit of modern smartphones.
Figure \ref{fig:energy_breakdown}
reveals the source of FlexSpec's energy efficiency. Cloud-Only approaches consume high energy (4.5 J/token) primarily due to the radio tail states (Communication Energy). By drafting $K$ tokens locally and sending them in a compressed burst, FlexSpec reduces radio active time significantly (Communication Energy drops to 1.2 J), yielding a 53\% total energy reduction compared to standard streaming.
Running large models fully on-device generates significant heat (higher than $80^{\circ}$C on Jetson), causing thermal throttling. FlexSpec shifts the heavy lifting to the cloud, maintaining a thermal profile (Low-Med) suitable for handheld usage.

\section{Conclusions}
In this paper, we proposed FlexSpec, a communication-efficient collaborative inference framework for evolving edge-cloud systems based on speculative decoding. FlexSpec was designed to address the scalability limitations of existing SD-based approaches under frequent cloud-side model updates and dynamic wireless conditions. To this end, we first introduced a shared-backbone architecture that enabled a single, static edge-side draft model to remain compatible with a family of evolving cloud-side target models, thereby eliminating repeated edge-side retraining and model synchronization. Then, we further developed a channel-aware adaptive speculation mechanism that dynamically adjusted the speculative draft length to balance end-to-end latency, communication overhead, and device energy consumption under time-varying network conditions.
Finally, we evaluated FlexSpec through extensive experiments, which demonstrated that FlexSpec consistently achieved superior performance over conventional baselines and provided an effective strategy for deploying LLMs in dynamic edge-cloud environments.

\bibliographystyle{IEEEtran}
\bibliography{reference}

\end{document}